# User involvement in the partner program of the educational social network

Ilya V. Osipov [1*], Anna Y. Prasikova [1], Alex A. Volinsky [2*]

[1] i2istudy.com, Krišjāṇa Barona Iela, 130 k-10, Rīga, Lv-1012, Latvija

[2] Department of Mechanical Engineering, University of South Florida, 4202 E. Fowler Ave., ENB118, Tampa FL 33620, USA

[*] Corresponding authors. Email: volinsky@usf.edu; Phone: (813) 974-5658; Fax: (813) 974-3539 (Alex A. Volinsky); Email: ilya@i2istudy.com (Ilya V. Osipov)

**Abstract**

The paper describes experiments to attract active online system users to the partner program. The objective is to grow the number of users by involving existing system users in viral mechanics. Several examples of user motivation are given, along with the specific interface implementations and viral mechanics. Viral K-factor was used as the metrics for the resulting system growth assessment. Specific examples show both positive and negative outcomes. Growth of the target system parameters is discussed.

## 1 Introduction

A partner program has been created for the online internet project dealing with educational social network, stimulating users to invite their friends for the user base growth. The project name is deliberately not revealed in this paper, since it's not written for the advertising purposes, but rather for sharing a valuable practical marketing experience.

The project is based on the freemium economic model, where for the marketing purposes the majority of the users can use the system for free, while the minority utilizes premium paid services (Lambrecht et al. 2014). For a typical freemium product, 2-10% of the users utilize paid services, contributing to the project economic success (Mäntymäki and Salo 2011; Seufert 2014). Free, non-paying users also contribute to the project by developing content, participating in ratings and interactions with other users. The main objective of the project, which plans to attract more users, is to motivate existing users to invite their friends to join the system by participating in the viral cycle (Osipov et al. 2014a, b; Lewis et al. 2013). Many successful online services utilize partner programs to stimulate users to invite their friends, based on the win-win principle, when both participants get bonuses (Gains 2014). Hooked on this idea, the authors have developed an interesting partner program, but were not sure how to increase the system user involvement in the partner program.

**2 Assessment methods**

2.1 K-factor as the system assessment parameter

K-factor (viral factor) is the ratio of the newly registered invited users and the total number of system users. Local K-factor is calculated for a given period of time, one week in this case. Without users leaving the system, the K-factor reflects the system viral growth (Ellis and Brown 2014). More information about the viral K-factor and it's relation with user departure and retention is described by Rigatuso (2014).

2.2 Statistical analysis and significance

All experiments have been conducted using a small online educational social network of 40,000 registered users with 1,000-5,000 daily active users. To test the effect of every utilized viral mechanics, statistical analysis was conducted. P-value for the K-factor before and after each viral mechanics implementation was calculated. The null hypothesis assumed that the K-factor did not change as a result of the implemented viral mechanics. In each case the p-value was less than 0.001-0.006, thus the authors have rejected the null hypothesis that the K-factor did not change as a result of the implemented viral mechanics.

**3 Experiments**

3.1 Initial online partner program

The first simplest idea was to place a visible button in the user interface "Invite a friend and earn

30 minutes!" (Figure 1). Time in minutes is the internal accounting unit in the system, based on the time bank principle (Marks 2012; Válek and Jašíková 2013). The user, interested in this offer, was directed to the custom interface, which contained ways for placing open referrals in social networks and sending personalized invitations (Osipov and Volinsky 2014; Ellis and Brown 2014). This approach simply did not work. While some users sent out invitations, the number of referred registered users was minimal, less than 1%. The local K-factor was about 1-1.5% at the time.

Developed educational social network is global and international in nature, designed for the world-wide use. People start to pay attention when asked for help, especially in the western world.

3.2 Method A: Asking for help

The authors have changed the message and used the following call to action: "Help our project!" (Figure 2). The interface button text was changed as an experiment to test if it would work. This button was displayed in the internal system interface, only seen by the registered users. Obviously the phrase "Help our project!" needed an explanation, and the pop-up window was added containing the following text. "Our project needs more users. We want to grow. Please invite your friends. You and your registered friend will get 30 minutes each as a bonus." (Figure 2). This pop-up window contained a green button at the bottom: "Help the project – invite" (Figure 2). The window was not modal and contained a closing cross in the right upper corner, giving the choice for the user to close it (Johnson et al 2012). Moreover, the window was closed by clicking anywhere outside the window. Curiously enough, 26% of the registered users clicked on the "Help our project!" button. However, when they saw the following pop-up window, they simply closed it. Nevertheless, the local K-factor increased by another 1%, meaning that twice more users were attracted as a result. The authors decided to continue the experiment further.

3.3 Method B: Denying the denial

The closing cross was removed from the right upper corner of the pop-up window. This window no longer disappeared after a click outside the window. Moreover, nothing happened with any kind of clicks. The additional button was added: "I would not like to help." Now the user saw two options to either help or deny help. The trick was that now the user could only explicitly deny help. There were no other options. This "denying the denial" method actually worked. The user was trapped in the psychological dilemma of either denying help, or helping. Majority of the users decided to stay on the "bright side" and 73% of the users clicked the help button. Certainly, most of them did not follow through with the invites when they saw the next interface window. However, the K-factor reached 4% as a result, which doubled the number of users involved viral mechanics. This is an impressive result for such a small interface change.

3.4 Method C: Perfectionist progress bar

After the users already invited their friends, or simply sent blind invitations, it was inappropriate to continue to ask them for help. However, since the user sent the invitation once, s/he is more likely to do it again. Instead of asking for help in the system interface, there was a progress bar added: "1 of 10. Invite 9 more friends for the VIP status" (Figure 4). Most people like to finish something they have already started (Fields and Cotton 2012). Thus, progress bars and survey completeness indicators work as good motivators to finish the work (Gnambs et al. 2010).

Unfortunately the authors could not test if this simple trick has increased the user involvement in the viral invitation process, since there was nothing to compare it with. The authors can arbitrarily say that there has been 1000% growth, since prior to using this viral mechanics, the users who already invited friends have not been involved. Nevertheless, the authors have been tracking the weekly K-factor, which reached 6.5%.

**4 Conclusions**

The paper showed that simple organization of the user interface along with the corresponding messages addressed to the users can play a crucial role in the system growth rate. These simple viral mechanics actually worked, backed by the local K-factor growth, verified by statistical analysis. In the long run the K-factor determines whether or not the project will be successful or fail completely.

**5 Acknowledgements**

The authors would like to thank the i2istudy.com team members for their dedicated efforts: Anna Prasikova, Vadim Grishin, Ilya Poletaev, Andrei Poltanov, Elena Bogdanova, Vildan Garifulin, Mihail Shagiev and Franziska Rinke. Valuable discussions about the data statistical analysis with Dr. Vladimir Kogan are greatly appreciated.

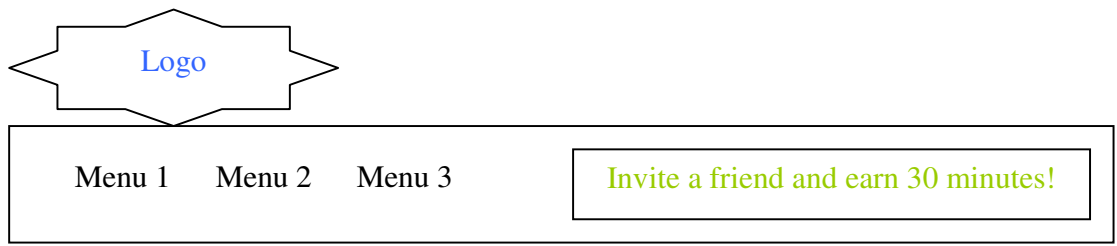

Figure 1. Schematics of the user interface with the green button to invite friends.

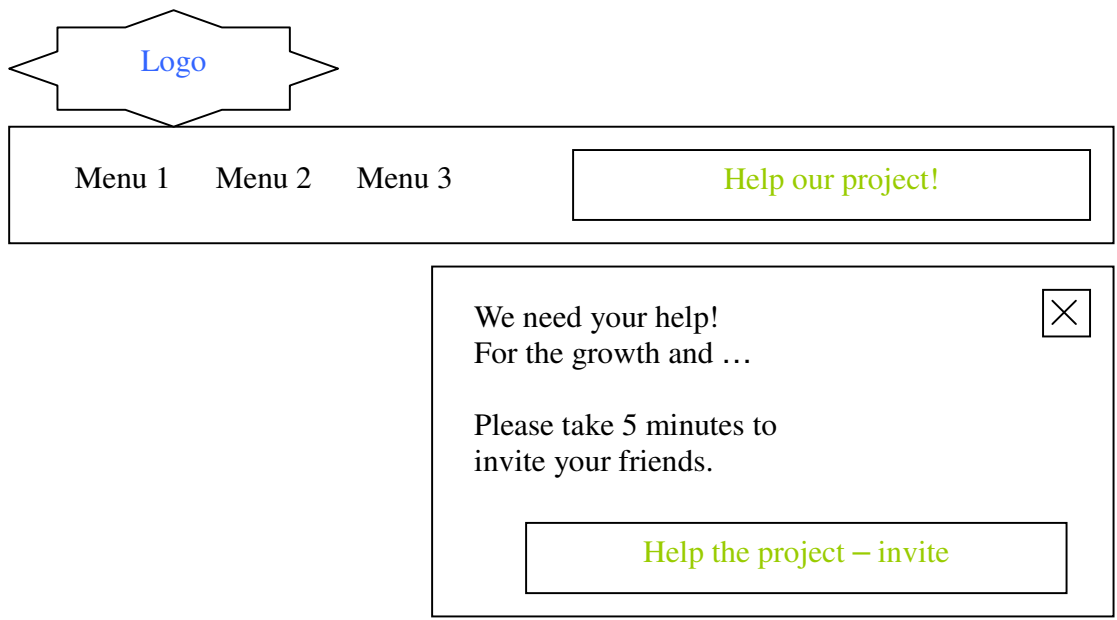

Figure 2. Schematics of the user dialogue pop-up window to invite friends with the cross in the upper right corner.

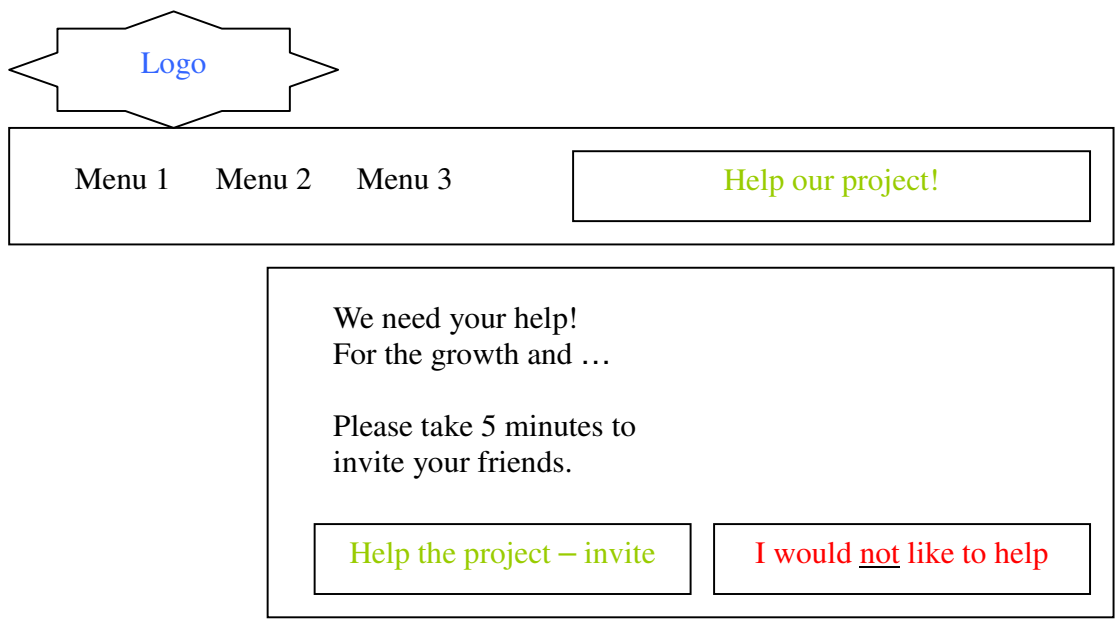

Figure 3. Schematics of the user dialogue pop-up window to invite friends with the red "I would not like to help" button and no cross in the upper right corner of the pop-up window.

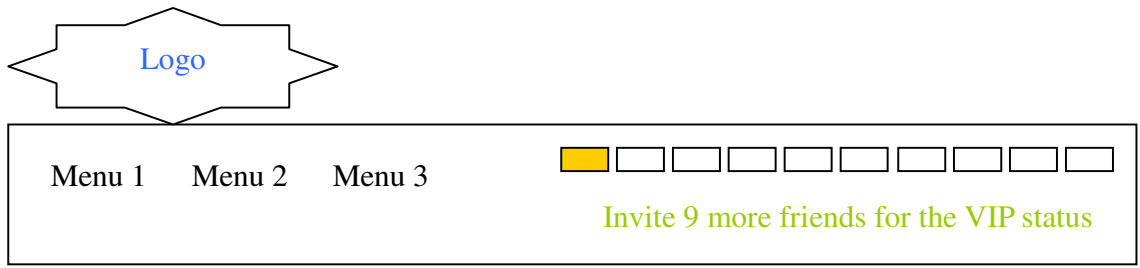

Figure 4. Schematics of the user interface with the bar showing the number registered friends needed to obtain the VIP status.